# First-principles investigation of the effect of substitution and surface adsorption on magnetostrictive properties of Fe-Ga alloys

Hui Wang and Ruqian Wu

*Department of Physics and Astronomy, University of California, Irvine, California, 92697, USA*

**ABSTRACT:** Materials with large magnetostriction are widely used in sensors, actuators, micro electromechanical systems, and energy-harvesters. Binary Fe-Ga alloys (Galfenol) are the most promising rare-earth-free candidates combining numerous advantages such as low saturation magnetic field (~200 Oe), excellent ductility and low cost, while further improving their performance is imperative for practical applications. Using density functional theory calculation, we report results of the effect of substituting small amount of additional elements X (eg. X = Ag, Pd and Cu) on magnetostriction of Fe-Ga alloys, and find that it may double the magnetostriction with a substitutional percentage of only 1.6%. Moreover, adsorbents with high chemical activity (eg. O or Os atoms) may affect the surface energy of different face-orientations of Fe-Ga alloys, indicating proper surface treatments are necessary to tune the alignment of Fe-Ga grains to achieve better performance. These results may be helpful to further optimize the magnetostrictive properties of Fe-Ga alloys for device applications.

Correspondence should be addressed to: wur@uci.edu

# 1. INTRODUCTION

Exploring novel magnetostrictive materials that can change their dimension with a small magnetic field is crucial for both fundamental research and technological exploitations [1, 2]. One of the most successful magnetostrictive materials hitherto is Terfenol-D ($Tb_{0.3}Dy_{0.7}Fe_2$) that shows giant magnetostriction up to 2000 ppm (parts per million), and has been widely used in different devices such as sensors, actuators, micro electromechanical systems, and energy-harvesters [3, 4], etc. However, their applications have been somehow limited due to the shortage of rare-earth supplies and mechanical brittleness. This inspired a new wave of interdisciplinary search for rare-earth free and ductile magnetostrictive materials. Fe-based materials especially $Fe_{1-x}Ga_x$ alloys (Galfenol with x~19%) are the most promising candidates as they exhibit excellent mechanical properties, large tetragonal magnetostrictive coefficient ($\lambda_{001}$~280 ppm), low saturation magnetic field (~200 Oe) and low cost [5-8]. Further development of these alloys for practical utilizations requires comprehensive understanding of the mechanism [6, 9-16] that governs the magnetostriction in transition metal alloys, from which we can develop viable approaches to further improve their magnetostrictive performance.

Recent experimental and theoretical studies suggest that the availability of non-bonding electronic states around the Fermi level is important for the initial quadratic increase of $\lambda_{001}$ of $Fe_{1-x}Ga_x$ alloys against x. Ga atoms avoid forming first neighbors in the Fe lattice and, as a result, the presence of each Ga atom effectively breaks 8 Fe-Fe bonds in Galfenol and hence many non-bonding Fe-d states are induced [12-15]. The dangling Fe-d states around the Fermi level allow strong spin-orbit coupling (SOC) interactions among them, and hence lead to a monotonic increase of the magnetoelastic coupling ($b_1$) with x up to x~15%. Meanwhile, the loss of Fe-Fe bonds reduces the tetragonal shear modulus ($c'$), from 60 GPa for the pure bulk Fe to about 10 GPa for $Fe_{81}Ga_{19}$ alloys. Since the tetragonal magnetostrictive coefficient $\lambda_{001}$ is simply the ratio of $b_1$ and $c'$ ($\lambda_{001}=2b_1/3c'$), it is apparent that both factors above contribute to the enhancement of $\lambda_{001}$.

In this paper, we report results of systematic density functional theory (DFT) calculations for the effect of substitution of several transition metal elements (eg. Ag, Pd and Cu) on the magnetostrictive properties of $Fe_{1-x}Ga_x$ alloys with x~19%. Interestingly, we found that a small substitutional amount of these elements may significantly enhance the magnetostriction of Galfenol, by a factor of >200%. Meanwhile, we also investigated the effect of different adsorbents (such as O atoms, Os atoms and $H_2S$ molecules) on the surface energies of Galfenol to provide useful guidance for the choice of chemical environment for the post-synthesis treatment of Galfenol samples, particularly for the preferential alignment of Fe-Ga grains along the (001) direction. These results provide new insights for the development of Galfenol with optimal performance in devices.

## 2. METHODOLOGY

Our DFT calculations were performed using the Vienna *ab initio* simulation package (VASP) [17, 18]. The exchange-correlation interactions were included using the spin-polarized generalized-gradient approximation (GGA) with the Perdew-Burke-Ernzerhof (PBE) functional [19]. We treated Fe-3d4s4p, Ga-4s4p, Cu-3d4s, Ag-4d5s, Pd-4d, H-1s, O-2s2p and S-3s3p as valence states and adopted the projector augmented wave method (PAW) to describe the valence-core interaction [20, 21]. 5×5×5 and 7×7×1 Monkhorst-Pack k-meshes [22] were used to sample the Brillouin zones of the bulk and surface models. The structures were fully relaxed with the criteria that require 1) the force acting on each atom is less than 0.01 eV/Å and 2) total energy convergence is better than $10^{-5}$ eV. The energy cutoff for the plane-wave expansion was set to 500 eV, which is sufficient for Fe-Ga alloys according to our previous studies.

To determine the magnetic anisotropic energy (MAE), we used the torque method [23] that calculate MAE as the expectation value of the angular derivative of the SOC Hamiltonian with respect to the polar angle θ of the spin moment, i.e., $\tau(\theta) = \frac{\partial E_{total}(\theta)}{\partial \theta} = \sum_{occ} \langle \psi_{i,k} | \frac{\partial H_{SO}}{\partial \theta} | \psi_{i,k} \rangle$. This approach has been successfully applied for studies of magnetic anisotropy of a variety of magnetic materials as well as for magnetostriction of many transition metal alloys [13, 24]. The bulk Fe-Ga alloys were simulated by a 4×4×4 supercell, which has 128 atoms in a cubic box. Their surfaces were mimicked by building up a slab model that consists of 9 atomic layers and a vacuum gap of about 12 Å thick to avoid the spurious interaction between periodic images. Different growth and annealing conditions were considered and simulated by varying the surface orientations and chemical adsorbents.

## 3. RESULTS AND DISCUSSION

### A. Substitutional effects on magnetostriction of Fe-Ga alloys

For the binary $Fe_{1-x}Ga_x$ alloys, the monotonic decrease of the tetragonal shear modulus continues up to x~25%, whereas the increase trend of the magneto-elastic coupling coefficient only sustains to x~15%. This causes the rapid drop of the magnetostriction after it reaches its maximum at x~19% [13]. Therefore, one needs to extend the uptrend of $b_1$ and remain relative small     beyond the critical Ga concentration. To this end, adding a small amount of the other elements is a promising way, and many elements including transition metals (e.g., Mn, Co, Ni, Cr, Zn) and metalloids (e.g, Ge, Si) have been used in previous studies [1, 5, 6, 14, 25-28]. Here, we choose the most stable $Fe_{79.7}Ga_{20.3}$ atomic structure obtained from our previous studies as the template and study the effect of Ag, Pd, Cu substitution on the magnetostrictive properties of Galfenol. The unit cell includes 102 Fe atoms and 26 Ga atoms, and we substitute two Ga atoms with X atoms (X = Ag, Pd, Cu) to form the $Fe_{79.7}Ga_{18.7}X_{1.6}$ ternary alloys. To figure out the preferential configuration of substitution, we change the separation between two X atoms from 2.45 Å, 2.91 Å, 4.09 Å to 10.02 Å, respectively, as marked by red in Fig. 1(a).

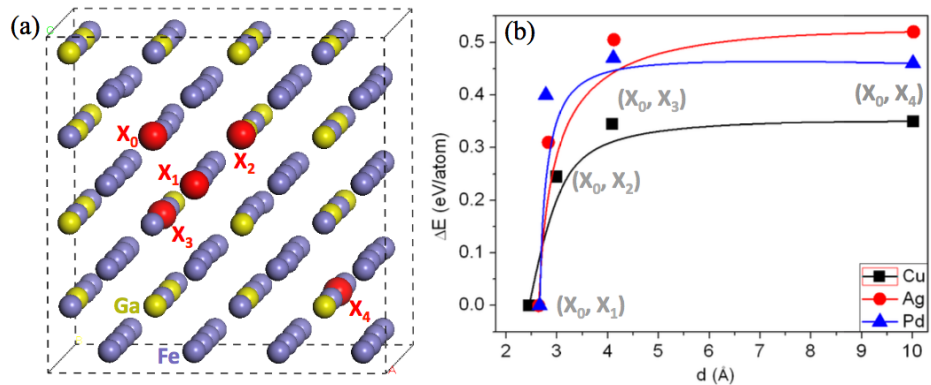

FIG. 1 (color online) (a) Schematic models for $Fe_{1-x}Ga_x$ alloys with a small mount of X elements at different distance varying from first $(X_0, X_1)$, second $(X_0, X_2)$, third nearest neighbors $(X_0, X_3)$ and even further $(X_0, X_4)$. The light blue, yellow and red represents Fe, Ga and X elements (X = Ag, Pd, Cu), respectively. (b) The relative energy difference of $Fe_{79.7}Ga_{18.7}X_{1.6}$ alloys as a function of the distance between two X atoms in Fe-Ga matrix as shown in (a), the fitted solid line are guided for your eyes.

We found that the total energy of $Fe_{79.7}Ga_{18.7}X_{1.6}$ ternary alloys (X = Ag, Pd, Cu)

remain almost constant when the distance of two X atoms ($d$) is larger than 4.09 Å, indicating the weak interaction between them at this region as shown in Fig. 1(b). However, the total energy decreases significantly up to 0.3~0.5 eV/X atom as two X atoms become the second or first nearest neighbors, due to their strong hybridization with each other. These results clearly indicate that the substitutional Ag, Pd and Cu elements prefer to stay together and may form clusters if the thermo-dynamical process is slow enough, in line with the poor solubility of these elements in the bcc Fe matrix [29]. Since clustering of these elements is detrimental to the magnetostriction according to our calculations, one may use fast cooling or quenching method to freeze the metastable distribution patterns of X elements in the $Fe_{79.7}Ga_{20.3}$ matrix to obtain high magnetostriction in $Fe_{79.7}Ga_{18.7}X_{1.6}$ ternary alloys.

Now we want to discuss the possibility of increasing tetragonal magnetostriction $\lambda_{001}$ with these substituents. Following to the rigid band model, we first analyze the dependence of MAE of a strained $Fe_{79.7}Ga_{20.3}$ lattice (±1% along the z-axis, while the lattice size in the lateral plane was adjusted according to the constant-volume mode: $\varepsilon_z = -2\varepsilon_x = -2\varepsilon_y$) on the total number of electrons in the supercell as shown in the lower panel of Fig. 2(a). Note that the Fermi level ($N_e = 1154$) touches the intersection of the two $MAE(N_e)$ curves, suggesting a weak magneto-elastic coupling (or small $b_1$) of $Fe_{79.7}Ga_{20.3}$ alloys. It is clear that the strain-induced MAE (or $b_1$) of $Fe_{79.7}Ga_{20.3}$ alloys can be further enhanced by either taking away (for positive $\lambda_{001}$) or adding (for negative $\lambda_{001}$) electrons to the unit cell, as shown by the green arrows in the lower panel of Fig. 2(a). Practically, this can be done through Ag, Pd, Cu or Ge, Si substitution for Ga atoms, respectively, assuming that they do not significantly affect the band structure of the Fe-Ga alloys near around Fermi level.

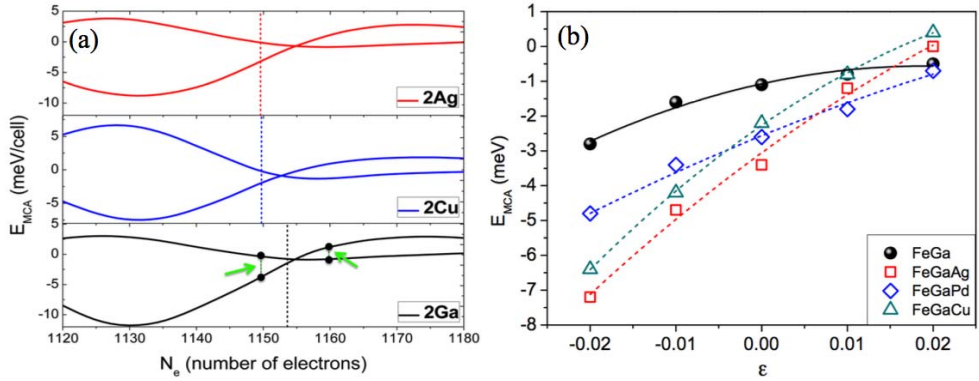

FIG. 2 (color online) (a) Calculated $E_{MCA}$ with $\varepsilon_z = \pm 1\%$ for $Fe_{79.7}Ga_{20.3}$ (black solid line), $Fe_{79.9}Ga_{18.7}Cu_{1.6}$ (blue solid line) and $Fe_{79.9}Ga_{18.7}Ag_{1.6}$ (red solid line) versus the number of valence electrons ($N_e$) in the unit cell. The vertical dash lines show corresponding positions of their actual $N_e$. The arrows indicate taking away or adding electrons to the unit cell. (b) Calculated strain dependent $E_{MCA}$ of FeGaX, where X represent Ag, Pd, Cu, respectively.

To verify our proposal through this method, we conduct DFT calculations for $Fe_{79.7}Ga_{18.7}X_{1.6}$ ternary alloys by replacing two Ga atoms in the 128-atom supercell with X = Ag, Pd, Cu atoms, respectively. Indeed, the trends of strain dependent MAE of these alloys are very similar, indicating that the uniform substitution of X for Ga rarely affect the band structure near the Fermi level. As depicted in the upper panels of Fig. 2(a), one can see that the Fermi level of $Fe_{79.7}Ga_{18.7}Ag_{1.6}$ and $Fe_{79.7}Ga_{18.7}Cu_{1.6}$ move to the left side by 4 electrons comparing with $Fe_{79.7}Ga_{20.3}$ since either Ag and Cu atom has two less electrons than Ga atom. As guided by the rigid band analysis, the strain induced MAE at the Fermi level (or magneto-elastic coupling coefficient $b_1$) are significantly larger than that of pristine $Fe_{79.7}Ga_{20.3}$ alloys as demonstrated in Fig 2(b). These results show the usefulness of manipulating number of electrons for the design of novel rare-earth-free magnetostrictive materials.

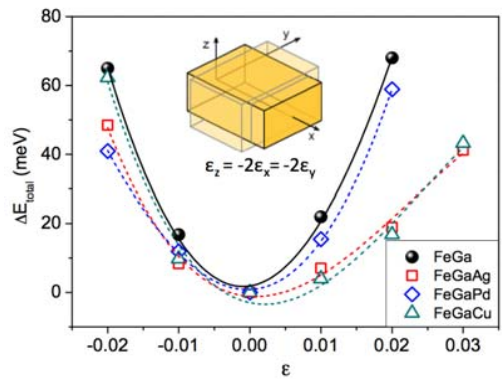

FIG. 3 (color online) Calculated strain dependent total energies of $Fe_{79.9}Ga_{18.7}X_{1.6}$ alloys, where X represent Ag, Pd, Cu, respectively. Inset demonstrates the applied strain under the condition of constant volume.

As we mentioned above, large magnetostriction relies on two main factors: strong magneto-elastic coupling coefficient $b_1$ and small tetragonal shear modulus $c'$. As known, $b_1$ and $c'$ are simply proportional to the slope of the MAE~ε line and the curvature of the total energy curve near ε=0%, respectively. From the strain induced changes of MAEs and total energies in Fig. 3, the calculated values of $b_1$ for $Fe_{79.7}Ga_{18.7}Ag_{1.6}$, $Fe_{79.7}Ga_{18.7}Pd_{1.6}$ and $Fe_{79.7}Ga_{18.7}Cu_{1.6}$ are ~17.5 $MJ/m^3$, 9.7 $MJ/m^3$ and 15.8 $MJ/m^3$, both are much larger (about 1.4~2.5 times) than that of the binary $Fe_{79.7}Ga_{20.3}$ alloy (~7.0 $MJ/m^3$). Meanwhile, the tetragonal shear modulus $c'$ for $Fe_{79.7}Ga_{18.7}Ag_{1.6}$, $Fe_{79.7}Ga_{18.7}Pd_{1.6}$ and $Fe_{79.7}Ga_{18.7}Cu_{1.6}$ ternary alloys are 8.6 GPa, 9.5 GPa and 9.7 GPa, respectively. In comparison, $c'$ of the pristine $Fe_{79.7}Ga_{20.3}$ alloy is close to 10.0 GPa. Therefore, the increase of magneto-elastic coupling constant $b_1$ is the main reason for the large enhancement of $\lambda_{001}$ in $Fe_{79.7}Ga_{18.7}X_{1.6}$ (X = Ag, Pd, Cu) ternary alloys.

**B. The effect of adsorbents on surface energies of Fe-Ga alloys**

It is known that the magnetostriction of Fe-Ga alloys is strongly anisotropic, namely, the tetragonal magnetostrictive coefficient, $\lambda_{001}$, can reach to about 280 ppm while its rhombohedral magnetostrictive coefficient, $\lambda_{111}$, is one order of magnitude smaller (± 20~30 ppm). Therefore, it is crucial to develop an approach that can align most Fe-Ga grains along the (001) direction in order to achieve an optimal performance. It is believed that the alignment of grains in Fe-Ga films depends mainly on the surface energies (denoted as "γ") of different facets, which can be controlled by tuning the chemical potential and using different surface adsorbents. Here, we consider the surface energy of a facet with adsorbents according to the following equation:

$$\gamma(N) = \frac{1}{2A}(E_{slab+M}(N) - N_{Fe}\mu_{Fe} - N_{Ga}\mu_{Ga} - N_M\mu_M) \qquad (1)$$

where $N_{Fe}$, $N_{Ga}$, and $N_M$ denote the numbers of atoms of Fe, Ga and adsorbent, respectively; $\mu_{Fe}$, $\mu_{Ga}$ and $\mu_M$ represent their corresponding chemical potentials. A is the surface area of the unit cell and the factor ½ accounts for the two surfaces in typical slab models. To allow direct comparison between different non-stoichiometric Fe-Ga facets,

we assume an equilibrium growth condition with a constraint of

(2)

where $\mu_{Fe13Ga3}$ is the chemical potential of the bulk $Fe_{13}Ga_3$ in the $D0_3$ structures, so we may use $\mu_{Ga}$ as a parameter to represent the different annealing condition.

Since the concentration of substituents that we discussed above is rather low, in principle they should not significantly alter the surface energies. For simplicity, we focus on the changes of surface energies of $Fe_{81.25}Ga_{18.75}$ alloy caused by different adsorbents such as oxygen atoms, heavy transition metal Os atoms and $H_2S$ molecules. According to the calculated total energies and comparing different adsorption sites, we find that O atoms prefer to take the atop-Ga site and Os atoms strongly bind to the bridge site of surface Ga atoms, with a binding energy of -4.25 eV/O atom and -6.23 eV/Os atom, respectively; while $H_2S$ molecule weakly adsorb on the atop-Ga site with a binding energy of -0.21 eV/$H_2S$ molecule. The most stable adsorption geometries and important bond distances are demonstrated in Fig. 4.

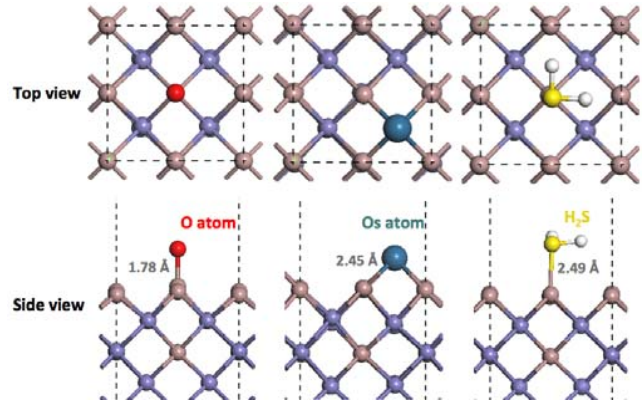

FIG. 4 (color online) The most preferential adsorption sites of O atom, Os atom and $H_2S$ molecule on Fe-Ga surface. Light blue, light red, yellow, white, red and cyan represent Fe, Ga, S, H, O, and Os, respectively.

Since hybridization between adsorbents and substrates may change the Fe-Ga surface energies, we then focus on calculating the surface energies of the (001), (110) and (111) facets at different Ga concentrations in the topmost layer with the presence of O atoms, Os atoms and $H_2S$ molecules. As we can see in Fig. 5, Ga atoms prefer to segregate toward the surface (at 100% Ga coverage) in the Ga rich condition ($\mu_{Ga} \to 0$) for all orientations. For example, the difference between surface energies of the Fe-terminated (0% Ga coverage) and the Ga-terminated (100% Ga coverage) surfaces is as large as 6.1 J/m$^2$ for Os atom/Fe-Ga(110) surface. In the Ga poor condition ($\mu_{Ga}$ < -3.0 eV), (001) and (110) surfaces with 75% Ga and 50% Ga coverage gradually become more stable. The critical condition occurs at $\mu_{Ga}$ = -2.6 eV for the O/Fe-Ga (001), $\mu_{Ga}$ = -3.2 eV for the Os/Fe-Ga (001) and $\mu_{Ga}$ = -3.0 eV for H$_2$S/Fe-Ga (001), respectively. It is interesting that Fe-Ga (111) surface prefer 100% Ga coverage in the entire range of chemical potential. We want to point out that the tendency of Ga segregation towards the surface self-stops as long as a monolayer Ga forms on the top according to our previous studies for clean Fe-Ga surface [30], and hence the Ga concentration in the interior region of Fe-Ga alloys is stable.

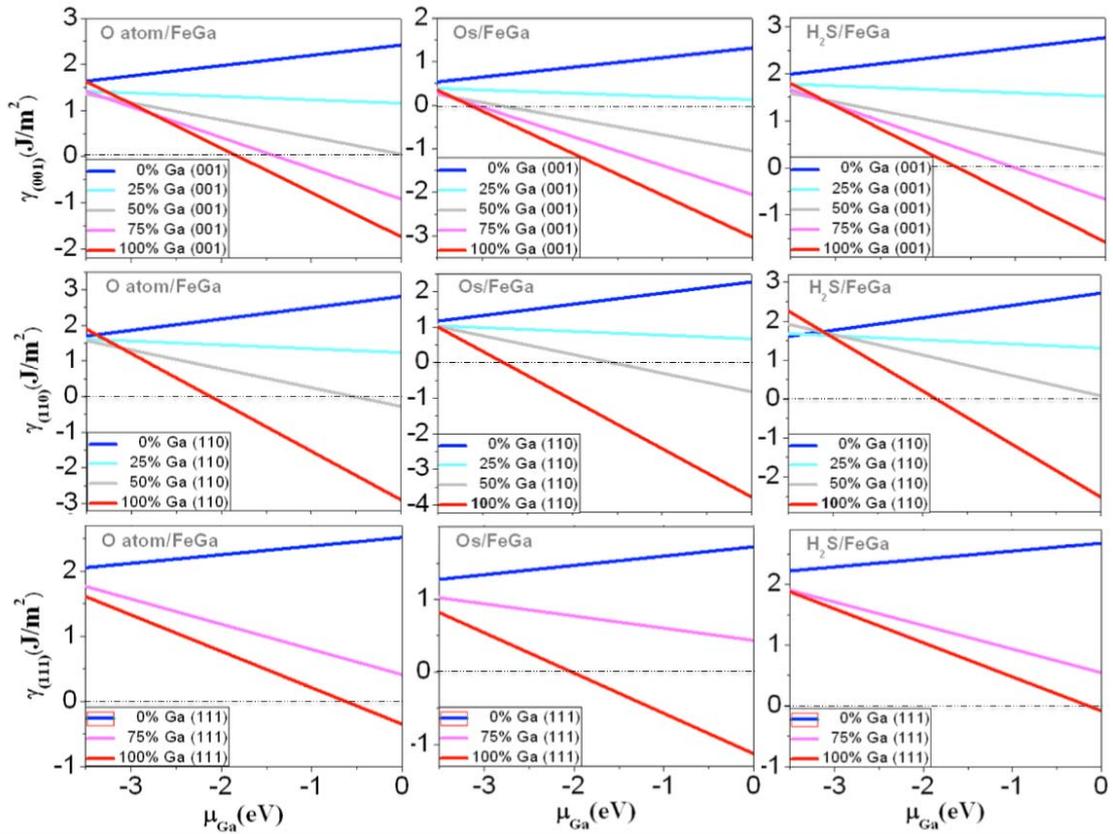

FIG. 5 (color online) The calculated surface energies for (001), (110) and (111) surfaces with different percentage of Ga coverage. Horizontal dash dot lines indicate zero energy. Left, middle and right panel represent O/Fe-Ga, Os/Fe-Ga and H$_2$S/Fe-Ga, respectively.

To highlight the effect of different adsorbents, we further compare surface energies of the most stable configurations of the three different orientations, i.e., 100% Ga (001), 100% Ga (110) and 100% Ga (111), as demonstrated in Fig. 6. With adsorbed O atoms,

Os atoms and $H_2S$ molecules, all (111) surfaces have much higher energies than their (001) and (110) counterparts so the formation of grains with the (111) orientation is largely suppressed, which is beneficial for the magnetostrictive performance of Fe-Ga films since $\lambda_{111}$ of Fe-Ga alloys is small and sometimes negative. It shows that the (110) surface is more stable in the Ga-rich condition ($\mu_{Ga} \to 0$) while (001) surface becomes more favorable in the Ga-poor condition ($\mu_{Ga} \to -3.0$ eV). As shown in Fig. 6(a), the crossover of surface energies between the (001) and (110) orientations with adsorbed O atoms appear at the left side of the normal Ga-poor condition ($\mu_{bulk-Ga} = -2.7$ eV), while for that with adsorbed Os atoms and $H_2S$ molecules it appear at the right side of Ga-poor condition [shown in Fig. 6(b) and (c)]. Among all adsorbents, adsorbed $H_2S$ doesn't affect the surface energies as compared with clean Fe-Ga surface, as demonstrated in Fig. 6(c) and (d). It is worth noting that adsorbed Os atoms push the intersection of Fe-Ga surface energies between the (001) and (110) orientations to the side of Ga-rich condition ($\mu_{Ga} = -1.8$ eV), which will be helpful for the formation of grains with the (001) orientation and maximize the magnetostrictive properties of Fe-Ga films. In the oxidation condition, one has to use a reservoir that binds to Ga atoms more tightly than the bulk Ga so as to create an environment for aligning Fe-Ga grains along the (001) direction. Nevertheless the energy difference between (001) and (110) surfaces is rather small in the Ga-poor end ($-3.0$ eV $< \mu_{Ga} < -2.0$ eV).

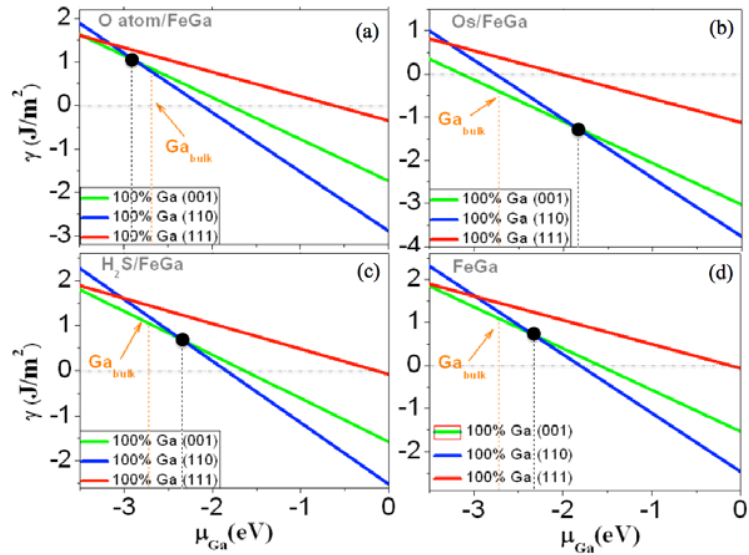

FIG. 6 (color online) Comparison of calculated Fe-Ga surface energies of the most stable configurations for

(001), (110) and (111) orientations with adsorbed (a) O atoms, (b) Os atoms, (c) H$_2$S molecules and (d) clean surface. The orange arrow indicates the chemical potential of orthorhombic bulk Ga; the black point represents the intersection of surface energies between (001) and (110) orientation.

In order to understand the role of different adsorbents, we calculated the projected density of states (PDOS) of O atoms, Os atoms and H$_2$S molecules adsorbed on Fe-Ga (001) surface with 100% Ga coverage. Adsorbed O atoms interact with underneath Ga atoms which are pulled up by ~0.43 Å comparing to their positions on the clean surface. As shown by the PDOS and charge redistribution in Fig. 7(a) and (b), O atoms strongly hybridize with Ga$_{surf}$ orbitals near the Fermi level, and the PDOS of surface Ga in O/Fe-Ga(001) is shifted to higher energy due to electron transfer from Ga to O. As a result, O adatoms significantly affect the surface energies of Fe-Ga alloys and the crossover of surface energies between the (001) and (110) orientations in O/Fe-Ga moves to the extreme Ga-poor condition. In contrast, adsorbed Os atoms transfer electrons from Os to Ga and interact with substrate significantly, pushing the crossover of Fe-Ga surface energies between the (001) and (110) orientations to Ga-rich condition. As also demonstrated in Fig. 7(c), H$_2$S adsorbed on Fe-Ga surface with a distance of ~2.5 Å and its electronic states mainly lie at -7.0 eV, far below Fermi level. The PDOS of surface Ga in H$_2$S/Fe-Ga(001) and clean Fe-Ga(001) remain almost unchanged below the Fermi level, indicating a rather weak interaction between H$_2$S and the Fe-Ga substrate. Therefore, the surface energies in H$_2$S/Fe-Ga and clean Fe-Ga are not much different. These results suggest that one may need to anneal Fe-Ga samples in the Ga poor condition and make proper surface treatments to promote most grain alignment along the (001) direction for better performance.

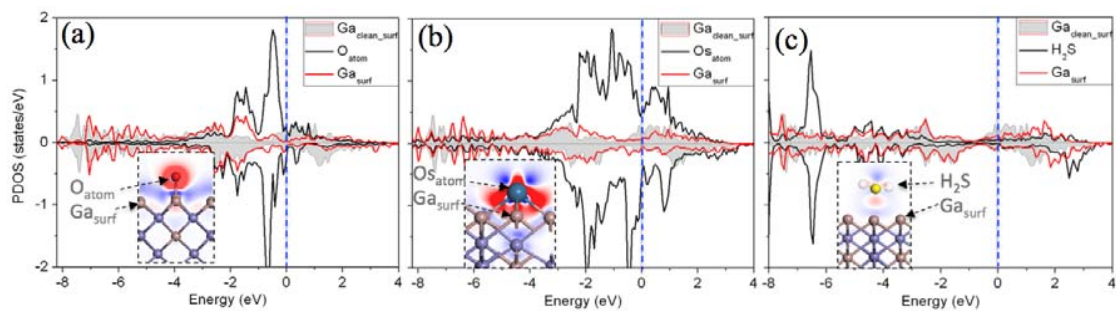

FIG. 7 (color online) The projected density of states (PDOS) of (a) O/Fe-Ga, (b) Os/Fe-Ga and (c) H$_2$S/Fe-Ga for (001) surface orientation with full Ga coverage, respectively. As a reference, shaded area demonstrates the PDOS of Ga atoms in clean Fe-Ga surface. Insets demonstrate the corresponding atomic configurations and charge redistribution between adsorbents and Fe-Ga substrate. Red and blue represent charge accumulation and depletion at 0.08 e/Å$^3$, respectively. Blue dash line indicates the Fermi energy.

## 4. CONCLUSIONS

In summary, we performed systematic DFT calculations to find possible ways for further improving the magnetostrictive properties of Fe$_{1-x}$Ga$_x$ alloys at x ~19%. Rigid band theory analysis suggests that this is realizable by substituting a small amount of Ag, Pd and Cu for Ga atoms in the Fe-Ga matrix, which is confirmed by DFT calculations with a large unit cell. Furthermore, the effect of different adsorbents on the surface energies of Fe-Ga alloys was also investigated, that may guide the design of growth and annealing conditions for the preferential (001) alignment of Fe-Ga grains in films. These results show the feasibility of engineering the magnetostrictive properties of transition metal alloys by tuning their electronic properties and surface environment for the optimal performance of these materials for device applications.

## 5. ACKNOWLEDGMENTS

We are grateful to Drs. A.E. Clark, M. Wun-Fogle, K.B. Hathaway, and A.B. Flatau for insightful discussions. This work was supported by the Office of Naval Research (Grant Nos: N00014-13-1-0445 and N00014-17-1-2905).